\documentclass[fleqn,twoside]{article}
\usepackage{espcrc2}
\usepackage{units}
\usepackage{amsmath}
\usepackage{amssymb}
\usepackage{subfig}

\usepackage{graphicx}
\usepackage[figuresright]{rotating}

\newcommand{\delm}[1]{\ensuremath{\Delta m^2_{ {#1} }}}
\newcommand{\mixsin}[1]{\ensuremath{\sin^2{(2\theta_{ {#1} })}}}

\newcommand{\nmue}{$\nu_\mu \rightarrow \nu_e \enskip$}

\newcommand{\numu}{\ensuremath{\nu_{\mu}}}

\newcommand{\numucc}{\ensuremath{\nu_{\mu}\!-\!\mathrm{CC}}} 
\newcommand{\nuecc}{\ensuremath{\nu_{\mathrm{e}}\!-\!\mathrm{CC}}} 
\newcommand{\nc}{\ensuremath{\mathrm{NC}}} 

\newcommand{\gnumi}{{\tt GNuMI}}
\newcommand{\gminos}{{\tt GMINOS}}
\newcommand{\gcalor}{{\tt GCALOR}}
\newcommand{\geant}{{\tt GEANT}}
\newcommand{\minos}{MINOS}
\newcommand{\numi}{NuMI}
\newcommand{\qe}{quasi-elastic}
\newcommand{\ngen}{{\tt NEUGEN}}

\newcommand{\ngthree}{{\tt NEUGEN-v3}}
\newcommand{\inuke}{{\tt INTRANUKE}}
\newcommand{\ranmod}{\ensuremath{\pi \rightarrow npnp}}

\title{Ramifications of intranuclear re-scattering in \minos}

\author{M.~Kordosky\address[UCL]{Department of Physics and Astronomy, University College London,\\  Gower Street, WC1E6BT London, United Kingdom}
}
       
\begin{document}

\begin{abstract}
\minos{} will measure the composition of a \numu{} beam at two locations, \unit[735]{km} apart, in an effort confirm the (atmospheric) neutrino oscillation hypothesis and measure the mixing parameters \delm{23} and \mixsin{23}. Oscillations will be manifested as a difference in the rate and energy spectrum of \numucc{} interactions measured in the two detectors. Because most interactions observed in \minos{} are inelastic, the \numu{} energy is reconstructed as the sum of the energy carried by the muon and that seen in the hadronic shower emanating from the struck nucleus. The latter is sensitive to uncertainties in the hadronisation process, chief among them those due to intranuclear re-scattering (i.e., final state interactions).  We discuss the simulation of intranuclear re-scattering currently used by \minos{} and its effect on quantities observable in the experiment.
\vspace{1pc}
\end{abstract}

\maketitle
\section{Introduction}
The Main Injector Neutrino Oscillation Search (\minos{}) is a long baseline, two-detector neutrino oscillation experiment that will use a muon neutrino beam produced by the Neutrinos at the Main Injector (\numi{}) facility at Fermi National Accelerator Laboratory (FNAL)~\cite{MINOS_Proposal,MINOS_TDR}.  The measurement will be conducted by two functionally identical detectors, located at two sites, the Near Detector (ND) at FNAL and the Far Detector (FD) in the Soudan Underground Laboratory in Minnesota.  The NuMI beamline~\cite{NuMI} and the \unit[735]{km} long-baseline will allow exploration of the \delm{23} and \mixsin{23} mixing parameters studied previously with atmospheric neutrinos~\cite{atmospheric-experiments} and by the K2K experiment~\cite{K2K}.  For \nmue transitions, \minos{} will probe neutrino mixing parameters beyond the current limits of the CHOOZ experiment~\cite{Apollonio:2002gd}.

The \minos{} detectors are tracking-sampling calorimeters, optimised to measure neutrino interactions in the energy range $\unit[1\lesssim E_{\nu}\lesssim 50]{GeV}$.  The active medium comprises \unit[4.1]{cm}-wide, \unit[1.0]{cm}-thick plastic scintillator strips arranged side by side into planes.  Each scintillator plane is encased within aluminum sheets to form a light-tight module and then mounted on a steel absorber plate. The detectors are composed of a series of these steel-scintillator planes hung vertically at a \unit[5.94]{cm} pitch with successive planes rotated by $90^{\circ}$ to measure the three dimensional event topology. Wavelength-shifting and clear optical fibers transport scintillation light from each strip to Hamamatsu multi-anode photomultiplier tubes which reside in light-tight boxes alongside the detector.  Both detectors are magnetized so as to measure muon charge-sign and momentum via curvature.

The \numu{} disappearance measurement will be done by using the event topology to identify interactions as \numucc{}, rather than \nc{}/\nuecc{}, then reconstructing the neutrino energy as the sum of the muon energy and shower energy: $E_{\nu} = E_{\mu}+E_{\mathrm{had}}$. The oscillation hypothesis will be tested by comparing the measured neutrino energy spectrum to the spectrum expected in the absence of oscillations.  The latter spectrum will be anchored to observations made with the Near detector. \minos{} endeavors, in the case of oscillations,  to measure \delm{23} and \mixsin{23} with an accuracy of better than 10\%.  The results depend upon a reliable knowledge of the event selection efficiency and the energy scale, both of which may be influenced by uncertainties in the hadronisation process.

\begin{figure*}[t]
  \begin{center}
    \subfloat[$\nu_{\mu}+A\rightarrow\mu^{-}+p+A^{\prime}$ with $E_{\nu}$={\unit[5.8]{GeV}}, y=0.07.]{

      \includegraphics[width=0.9\columnwidth,viewport=0 0 255 210,clip=true]{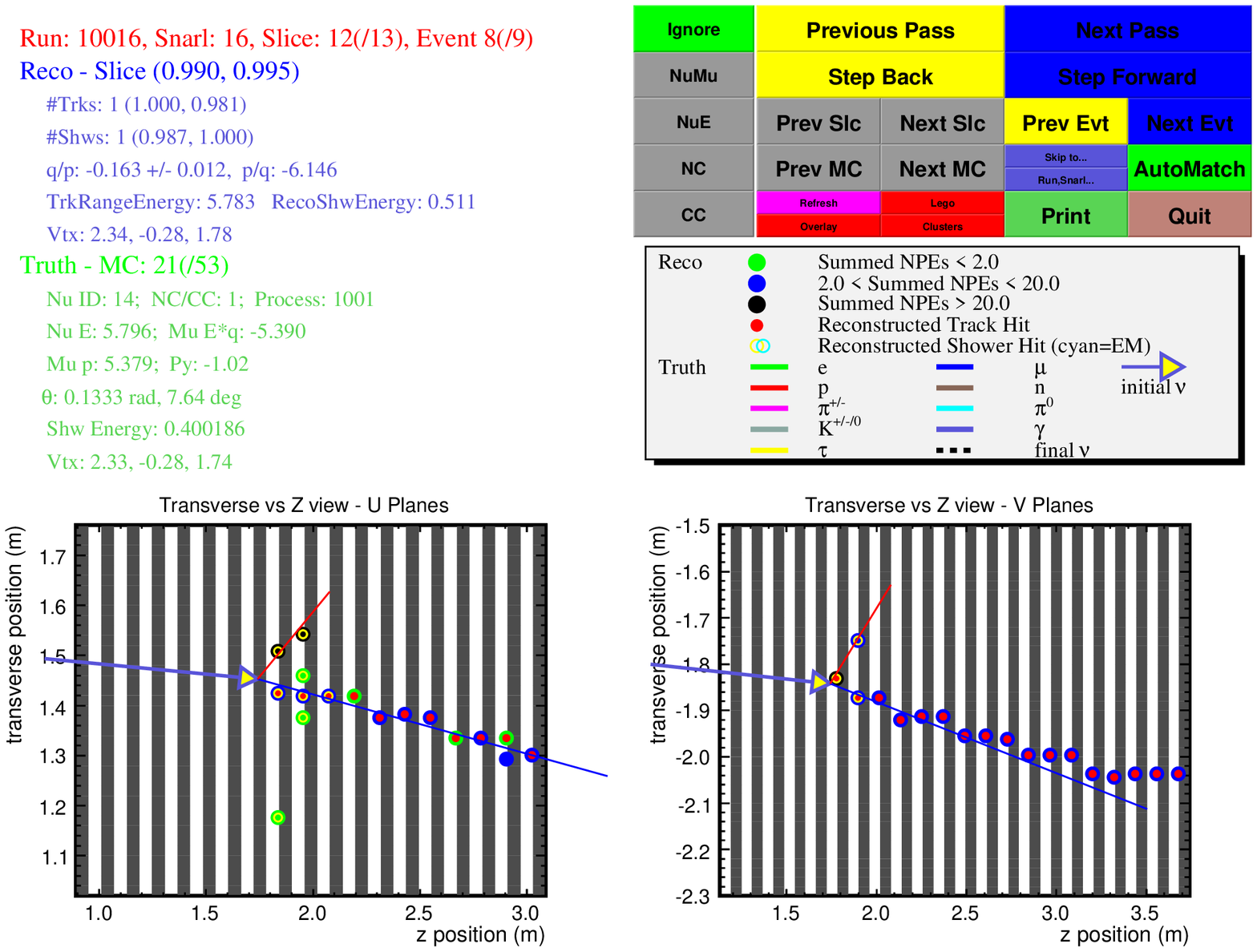}%
     \label{fig:qe_event}%
    }
    \qquad
    \subfloat[$\nu_{\mu}+~A \rightarrow \mu^{-} + \Delta^{++} + A^{\prime}$ with $E_{\nu}$={\unit[4.1]{GeV}}, y=0.25.]{
      \includegraphics[width=0.9\columnwidth,viewport=0 0 255 210,clip=true]{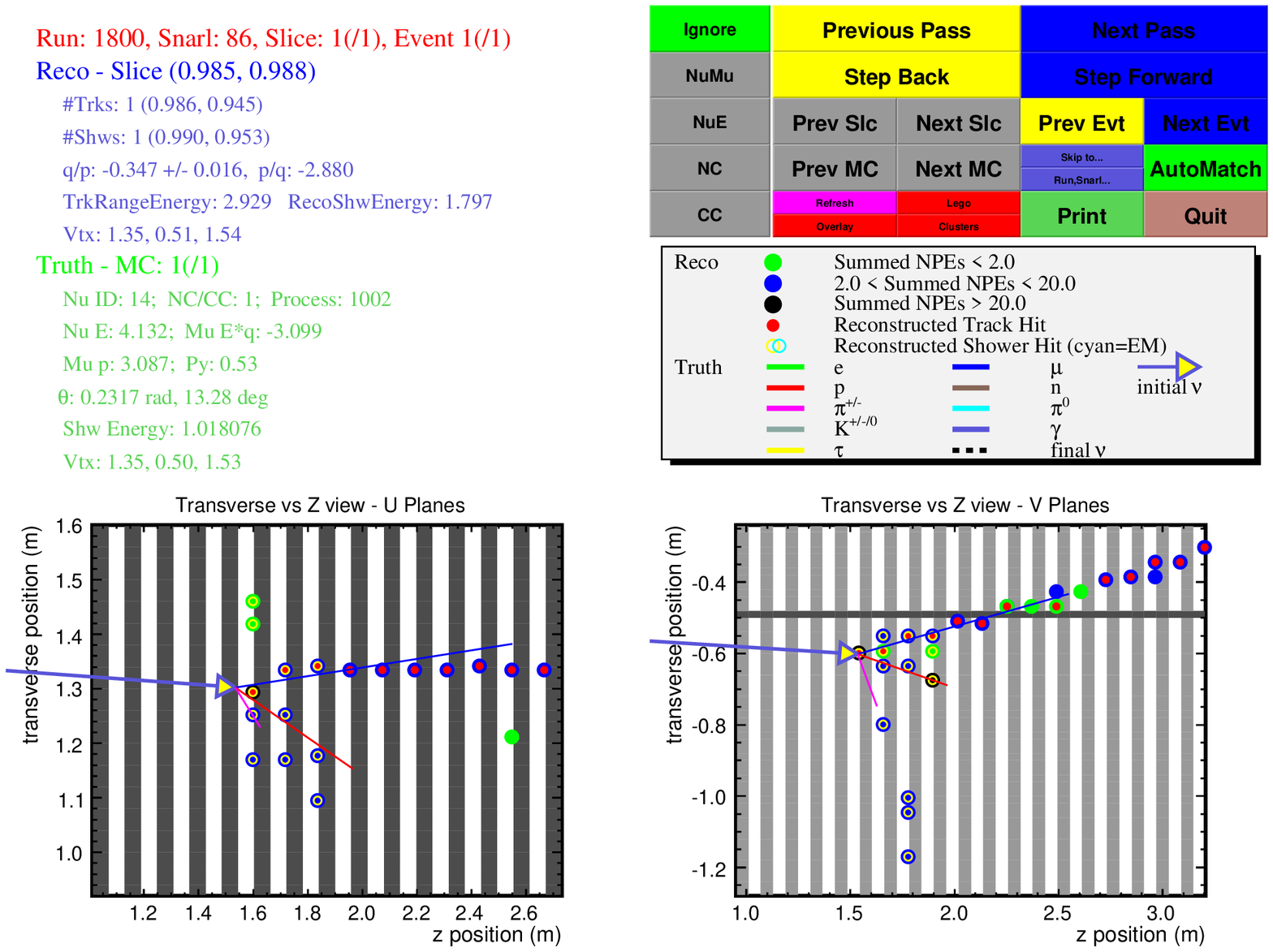}%
     \label{fig:res_event}%
    }
  \end{center}
  \caption{Two simulated events in the \minos{} Near Detector, zoomed to show the vertex region. The large arrow marks the direction of the incoming neutrino, lines mark the direction of the outgoing muon (blue), protons (red) and pions (pink), and dots show hits recorded by the detector. The dark vertical bands denote the iron absorber plates (not to scale). \minos{} measures two orthogonal ``views'' of each event of which only one is shown here.  }
\label{fig:events}
\end{figure*}

\section{Monte Carlo Simulation}

We will show results from a full Monte Carlo simulation of the \minos{} experiment. The simulation consists of four parts: (1) a neutrino target and beamline simulation (\gnumi{}), (2) the neutrino interaction generator \ngen{}~\cite{Gallagher:2002sf}, (3) a full detector \geant{}-based physics simulation known as \gminos{} and (4) a detailed active detector response simulation. The detector physics and response simulation has been benchmarked against cosmic ray and beam muons measured in the Near and Far detectors as well as series of testbeam measurements collected with a scaled down version of the \minos{} detectors in the CERN PS East Hall. The test beam measurements are used to fix the energy scale~\cite{Hartnell:2005uq} (via muon range) and validate the simulation of electromagnetic~\cite{Vahle:2004mp} and hadronic showers~\cite{Kordosky:2004mn}. The hadronic shower code \gcalor{} was found to be in good agreement with the data and is used in our simulations.

Neutrino interactions are modelled by \ngthree{}. \ngen{} was originally written to simulate neutrino backgrounds in the Soudan--2 proton decay experiment. Since that time it has undergone substantial modification to upgrade it for use at higher energies and has been adopted by \minos{}. \ngen{} includes a Rein-Seghal based treatment of neutrino induced resonance production, CC and NC coherent pion production and a modified leading order DIS model extended to improve the treatment in the transition region between DIS and resonant production~\cite{Bodek:2004ea}. KNO scaling is used to model the final state multiplicity in the DIS regime. 

Two simulated events in the are shown in Fig~\ref{fig:events}. The detector has a relatively fine segmentation for a neutrino calorimeter but was optimised to maximise the event rate in the FD, equalise performance of the two detectors, discriminate between $\nu$/$\bar{\nu}$, and cover a wide range of neutrino energies. The detector is not able to measure details of the final state in most interactions. Figure~\ref{fig:qe_event} is notable in that it contains a short but distinctive track-stub corresponding to a $\sim\unit[1]{GeV/c}$ proton. This proton is near the lower practical limit for identification in \qe{} events.

\ngen{} also includes an intranuclear re-scattering package known as \inuke{}. The code is anchored to a comparison of final states in $\nu + d$ and $\nu + \mathrm{Ne}$ interactions as measured in the BEBC and ANL--\unit[12]{ft} bubble chambers~\cite{Merenyi:1992gf}. The library includes a treatment of pion elastic and inelastic scattering, single charge exchange and absorption in a cascade simulation of the final state.  The relative probabilities for these processes are approximately 35:50:10:5 (40:30:7:23) at a pion kinetic energy of \unit[1]{GeV} (\unit[250]{MeV}). The formation zone concept is included by suppressing interactions that would have occurred before the pion has travelled a distance $l=\tau_{0}p/m$ $\tau_{0}=\unit[0.52]{fm/c}$~\cite{SKAT,flen}. The simulation includes a treatment of pion absorption inspired by~\cite{Ransome:2005vb}. Absorbed pions transfer their energy to a $npnp$ cluster and the individual nucleons are then tracked in \geant.
\section{Results}
\begin{figure*}[]
  \begin{center}
    \begin{minipage}[t]{\columnwidth}
    \includegraphics[width=0.9\columnwidth]{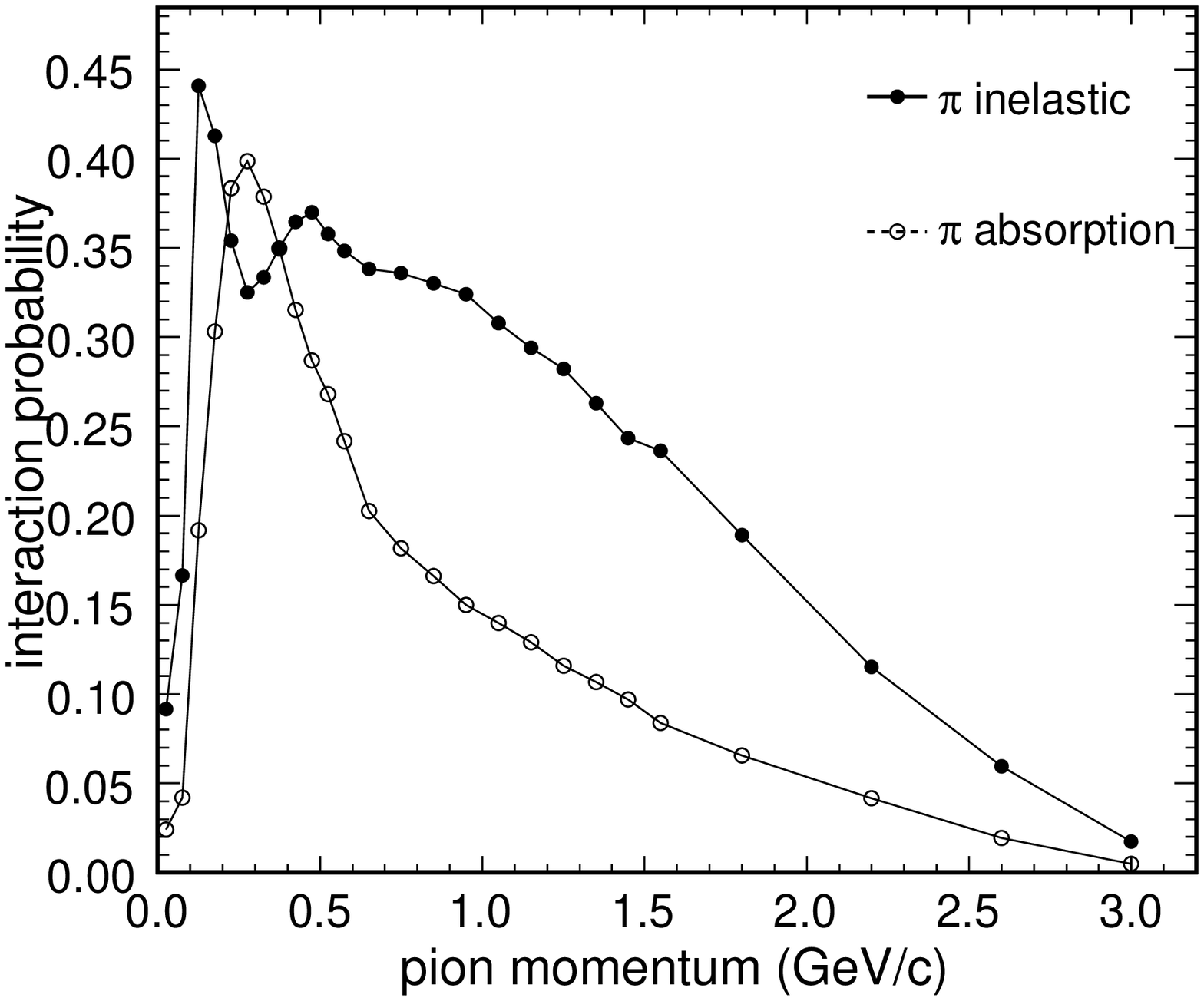}%
     \caption{The probability, according to \ngen{}, that a  pion produced in a neutrino interaction on $^{56}\mathrm{Fe}$ undergoes at least one inelastic interaction (full circles) or is absorbed (open circles) before exiting the nucleus.\label{fig:iprob}}   
     \end{minipage}
    \qquad
    \begin{minipage}[t]{\columnwidth}
    \includegraphics[width=0.9\columnwidth]{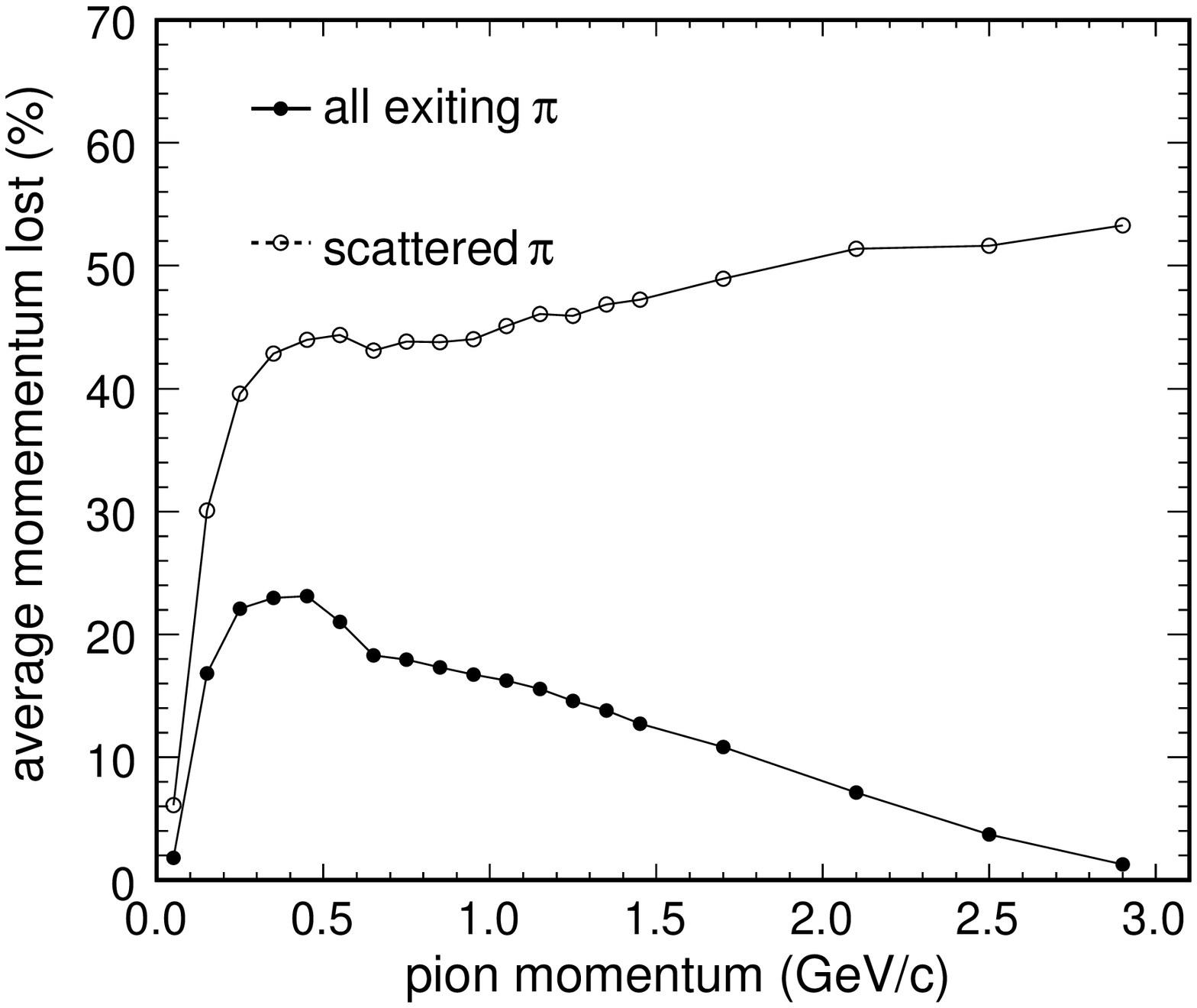}%
     \caption{The average momentum lost, due to intranuclear re-scattering, by pions produced in neutrino interactions on $^{56}\mathrm{Fe}$. The curves are the result of \ngen{} simulations. The full circles correspond to all pions exiting the nucleus while the open circles correspond to only those pions that scattered while exiting. \label{fig:ploss}}   
     \end{minipage}

    \begin{minipage}[t]{\columnwidth}
    \includegraphics[width=0.9\columnwidth]{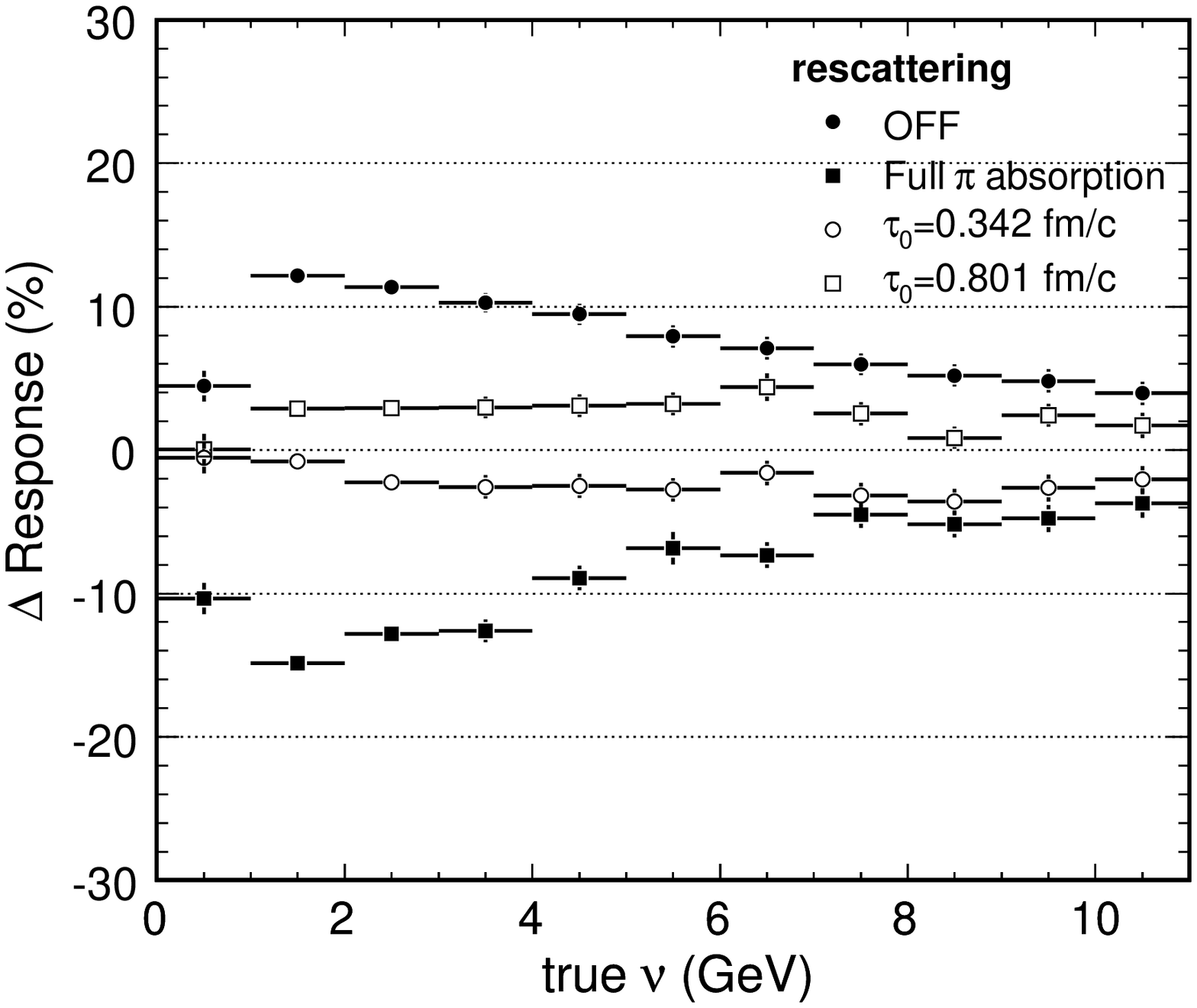}%
     
     \caption{The variation in the calorimeter's response as a function of the energy transfer $\nu$ and under different final state interaction scenarios. Zero on the vertical scale corresponds to the \ranmod{} model of pion absorption and $\tau_{0}$=\unit[0.52]{fm/c}.\label{fig:delta_eshw}}   
     \end{minipage}
    \qquad    
    \begin{minipage}[t]{\columnwidth}
    \includegraphics[width=0.9\columnwidth]{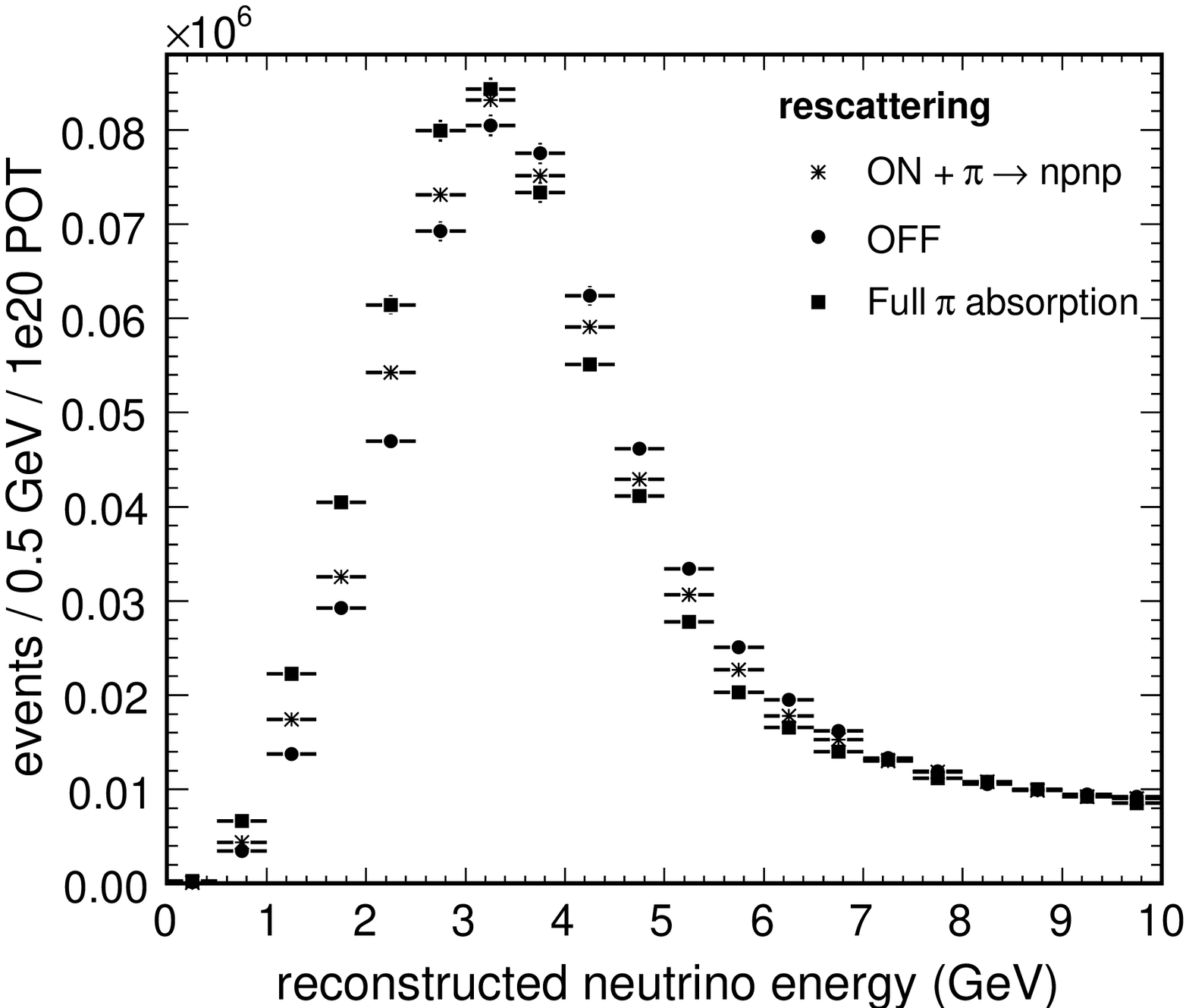}%
     \caption{The reconstructed energy spectrum of events selected as \numucc{} under three different final state interactions scenarios. The events were generated with \ngen{} and simulated in a full \geant{} simulation of the \minos{} Near Detector. The underlying spectrum is that expected from the \numi{} low-energy beam configuration.\label{fig:enuspec}}   
     \end{minipage}
  \end{center}
\end{figure*}

Figure~\ref{fig:iprob} shows the probability for absorption and inelastic scattering as a function of the initial pion momentum. Some familiar features, such as the $\Delta$ peak can be clearly seen. The decrease in the scattering probability with pion energy is a consequence of the formation zone.  The effective absorption probability is significantly larger than the relative probability stated above. This is caused by the fact that the assumed nuclear radius is much larger than than the mean free path, making multiple interactions common. Also, more importantly, pion absorption is a ``one way street'' in the sense that the cascade ends and other processes are suppressed. 

Pions lose a large fraction of their momentum in the intranuclear re-scattering process. Figure~\ref{fig:ploss} shows the average loss as a function of the pion momentum for all pions exiting the struck nucleus and separately for only those pions that underwent an inelastic interaction. The underlying distribution in the latter case is quite broad. The simulation assumes that energy lost in inelastic interactions is predominantly transferred to heavy remnants of the target nucleus which are expected to range out in the iron absorber, remaining invisible to the experiment.

It is clear that low energy, fully active, neutrino scattering experiments seeking to measure exclusive final states will be rather sensitive to re-scattering. However, based on the rather high interaction probability and significant energy loss, one should also expect observable consequences even in a dense neutrino calorimeter like \minos{}. This is indeed the case. Figure~\ref{fig:delta_eshw} shows the change in the measured vertex shower energy as a function of the energy transfer $\nu$. The origin of the vertical axis corresponds to the default configuration employed in our simulations. We observe a substantial change in the measured shower energy under different re-scattering scenarios. 

The difference between the default simulation and that without any re-scattering indicates that as much as 12\% of the initial shower is expected to go unobserved. Neutrino oscillation experiments that directly use $E_{\nu}$ depend upon reconstructing the energy (see Fig~\ref{fig:enuspec}) in an unbiased fashion or, at the very least, applying a correction factor when extracting the oscillation parameters. In this regard \minos{} is in a somewhat better position than its predecessors since the measured spectrum in the Near Detector provides a strong constraint when testing for oscillations and extracting the mixing parameters. However utilising this data requires a reasonable understanding of the accuracy of the rescattering formulation in order to disentagle its effects from those caused by uncertainties in the neutrino flux and cross-sections.


In Fig~\ref{fig:delta_eshw} ``Full $\pi$ absorption'' corresponds to absorption with the same probability as in Fig~\ref{fig:iprob} but in which the $\pi$ energy is not redistributed to an $npnp$ cluster. It is clear that the details of the pion absorption process are very important and that the shower energy measurement is sensitive to nucleons with a kinetic energy of only a few hundred MeV. Figure~\ref{fig:delta_eshw} also shows that the effect of final state interactions is not confined only to low shower energies. This is because even in large hadronic showers the energy is carried by multiple, lower energy particles. We have explored this effect by simulating with formation times  $\tau_{0}=\unit[0.34]{fm/c}$ and \unit[0.80]{fm/c}. This is a large variation but only results in a 2-3\% modification of the detector response.

Since \minos{} must use the event topology to remove the NC background we have studied the effect that re-scattering has on the efficiency and purity of the sample.  The selection procedure chooses events as \numucc{} based on the presence of a track with a length consistent with a muon having a momentum larger than \unit[500]{MeV/c}, one or or zero showers near the track vertex,  a good fit to the curvature in the magnetic field and a track pulse-height consistent with one MIP per plane crossing. We find that with this selection the \numucc{} efficiency and the level of NC background varies no more than a 1-2\% over the energy range $\unit[1<E_{\nu}<20]{GeV}$. Such small variations can be constrained using data from the Near detector. 

Finally, a more interesting situation occurs when we attempt to select a quasi-enriched sub-sample of events according to a procedure that uses the reconstructed invariant mass, the pulse-height not assigned to the track, the fraction of hits with a pulse-height larger than \unit[20]{PE} and the number of hits assigned to secondary tracks found via a Hough transform. Because baryon interactions are not simulated, the efficiency of this selection is independent of the final state interaction scenario.  The purity of the selection, however, is not.  Without re-scattering the purity of our sample integrated over \unit[0-10]{GeV} was 80\%, under the \ranmod{} scenario it decreased to 73\% and with full $\pi$ absorption only 63\% . This behaviour is broadly in line with expectations as our procedure tends to select events with low vertex activity as quasi-elastic.

\section{Conclusions}
We have explored the ramifications of intranuclear re-scattering in \minos{} through Monte Carlo studies using the \ngen{} event generator coupled to a \geant{} based detector simulation. The results of the study indicate that the reconstructed shower energy and topology depends rather strongly on the amount of energy lost in the re-scattering process. Though the efficiency for selecting an inclusive \numucc{} sample and the resulting sample purity are not strongly dependent on the re-scattering model, we find that the purity of a quasi-elastic sub-sample varies by $\sim 10\%$ when we compare simulations with and without rescattering. We expect that efforts to select other exclusive final states (such as coherent $\pi$ production) or to discriminate $\nu_{e}$ from NC, will be similarly afflicted.

The model employed here is under active development intended to widen its scope and increase its sophistication. The addition of baryon re-scattering seems quite important as does a more detailed treatment of the energy lost in inelastic collisions. Though the model is anchored to $\nu+d$ and $\nu+\mathrm{Ne}$ data, further benchmarking, either against additional data or theoretical predictions, would be very valuable. For example, the author would be interested in independent predictions of Figures~\ref{fig:iprob} and \ref{fig:ploss}. 

\section*{Acknowledgements}
The author would like to thank H.~Gallagher for his many helpful comments and technical assistance with the work presented here. The author was pleased to receive much helpful advice from W.~Mann and J.~Morfin and is grateful to R.~Hatcher and N.~Tagg for their assistance with the detector simulation. The quasi-elastic selection discussed here was implemented by M.~Dorman.

\bibliography{nuint05}

\begin{thebibliography}{10}
\bibitem{flen}
The formation length had been modified from the one recommended in~\cite{SKAT}
  to account for an effective nuclear radius used in the simulation.
\end{thebibliography}
\bibliographystyle{elsart-num}

\end{document}